\documentclass[preprint,12pt]{elsarticle}
\usepackage{graphicx}
\usepackage{epsfig}
\usepackage{mathtools}
\usepackage{amsmath,amssymb,amsfonts}
\usepackage{color}
\usepackage{enumerate}
\usepackage[T1]{fontenc}
\usepackage[utf8]{inputenc}
\usepackage[english]{babel}
\usepackage{lineno}

\date{}

\journal{Computer Methods and Programs in Biomedicine}

\begin{document}

\begin{frontmatter}

\title{Impaired coronary blood flow at higher heart rates during atrial fibrillation: investigation via multiscale modelling}

\author{S. Scarsoglio$\sharp ^*$, C. Gallo$\sharp$, A. Saglietto$\natural$, L. Ridolfi$\flat$, and M. Anselmino$\natural$\\
\it $\sharp$ Department of Mechanical and Aerospace Engineering\\
\it Politecnico di Torino, Torino, Italy\\
\it $\natural$ Division of Cardiology, "Città della Salute e della Scienza di Torino" Hospital\\
\it Department of Medical Sciences, University of Turin, Italy\\
\it $\flat$ Department of Environment, Land and Infrastructure Engineering\\
\it Politecnico di Torino, Torino, Italy}

\address{* Corresponding author: stefania.scarsoglio@polito.it\\ Department of Mechanical and Aerospace Engineering, Politecnico di Torino\\ Corso Duca degli Abruzzi 24, 10129, Torino, Italy. Phone: +39-0110906821.}

\begin{abstract}

\noindent \textbf{Background.} Different mechanisms have been proposed to relate atrial fibrillation (AF) and coronary flow impairment, even in absence of relevant coronary artery disease (CAD). However, the underlying hemodynamics remains unclear. Aim of the present work is to computationally explore whether and to what extent ventricular rate during AF affects the coronary perfusion.

\noindent \textbf{Methods.} AF is simulated at different ventricular rates (50, 70, 90, 110, 130 bpm) through a 0D-1D multiscale validated model, which combines the left heart-arterial tree together with the coronary circulation. Artificially-built RR stochastic extraction mimics the \emph{in vivo} beating features. All the hemodynamic parameters computed are based on the left anterior descending (LAD) artery and account for the waveform, amplitude and perfusion of the coronary blood flow.

\noindent \textbf{Results.} Alterations of the coronary hemodynamics are found to be associated either to the heart rate increase, which strongly modifies waveform and amplitude of the LAD flow rate, and to the beat-to-beat variability. The latter is overall amplified in the coronary circulation as HR grows, even though the input RR variability is kept constant at all HRs.

\noindent \textbf{Conclusions.} Higher ventricular rate during AF exerts an overall coronary blood flow impairment and imbalance of the myocardial oxygen supply-demand ratio. The combined increase of heart rate and higher AF-induced hemodynamic variability lead to a coronary perfusion impairment exceeding 90-110 bpm in AF. Moreover, it is found that coronary perfusion pressure (CPP) is no longer a good measure of the myocardial perfusion for HR higher than 90 bpm.

\end{abstract}

\begin{keyword}
atrial fibrillation \sep coronary blood flow \sep cardiovascular multiscale modelling \sep heart rate \sep computational hemodynamics
\end{keyword}

\end{frontmatter}


\section{Introduction}

Atrial fibrillation (AF) is the most common self-sustained arrythmia, leading to faster and irregular beating \cite{Kirchhof2016}. It is associated with increased morbidity and mortality \cite{Kannel2008} and is becoming a major public health burden in developed and developing countries \cite{Chugh2014}. AF induces several disabling symptoms such as palpitations, fall in blood pressure, decreased exercise tolerance, pulmonary congestion \cite{Fuster2006}. Beside these, AF patients may sometimes experience angina-like chest pain and symptoms attributable to alterations of the myocardial perfusion and transient myocardial ischemia, even in the absence of significant coronary artery disease (CAD) \cite{Kochiadakis2012,Wichter2010}. The impact of the beat-to-beat AF variability on the vascular structure has been recently investigated, with both clinical and computational approaches \cite{Olbers2018,Scarsoglio2018}. However, many hemodynamic questions still have to be disentangled. For example, regarding the coronary circulation - which is ruled by the diastolic phase - it is so far unknown which of the myocardial layers is mainly and first affected by AF rhythm. Moreover, it is important to understand whether repeated and continuous exposure to AF irregular beating can damage in the long-term the proper coronary functioning \cite{Kochiadakis2012}.

The reciprocal implications between AF and coronary perfusion impairment have long been discussed \cite{Wichter2010}. Although clear causal relations are still lacking, recent clinical studies suggest that AF promotes an impairment of the coronary flow \cite{Luo2014} and a mismatch of coronary blood flow and myocardial metabolic demand \cite{Kochiadakis2012}. Even in absence of significant CAD, AF can exacerbate the subendocardium perfusion during diastole. Several mechanisms - such as reduced aortic pressure related to short RR intervals, coronary vasoconstriction, and reduced coronary blood flow - have been proposed to explain how irregular ventricular rhythm can exert negative effects on the coronary hemodynamics. However, the precise mechanisms through which AF impacts the coronary circulation still remain to be explored \cite{Kochiadakis2002}.

Since CAD and AF frequently coexist and share several risk factors, clinical literature mostly considers the concomitant presence of both pathologies, e.g. \cite{Michniewicz2018,Martin2017,Kralev2011}. On one hand, this helps in shedding light on their interaction but, on the other hand, does not help in capturing the net impact of AF on coronary hemodynamics. If the role of irregular ventricular rhythm in AF is still unexplored to date, another open question is related to the impact of ventricular rate on the coronary perfusion. Even though lenient (resting HR <110 bpm) and strict (resting HR <80 bpm) rate control strategies were found not to differ in terms of mid-term cardiovascular outcomes \cite{VanGelder2010}, definitive evidence is still missing \cite{Wyse2011} and specific data referred to coronary flow impairment are overall scarce.

We therefore propose a computational approach, based on a validated multiscale model, to investigate to what extent coronary perfusion is impaired in AF and how this relates to the ventricular rate. Modeling aspects of the 0D-1D multiscale approach are recalled in Section 2, together with the stochastic RR beating extraction and the definition of the hemodynamic parameters. Results are presented and discussed in Sections 3 and 4, focusing on waveform and amplitude variations of left anterior descending (LAD) artery, as well as on coronary blood flow perfusion. Limiting aspects are reported in Section 5, which precedes the Conclusions (Section 6).

\section{Methods}

The proposed computational approach can be sketched as a three-steps algorithm: (a) RR beating extraction, (b) 0D-1D cardiovascular modelling resolution, (c) coronary perfusion evaluation. The three phases are described in the following subsections and schematically represented as overall flow chart of the stochastic modelling approach in Fig. 1.

\begin{figure}[h!]
	
	\includegraphics[width=1.05\columnwidth, trim=10 5 10 10, clip=true]{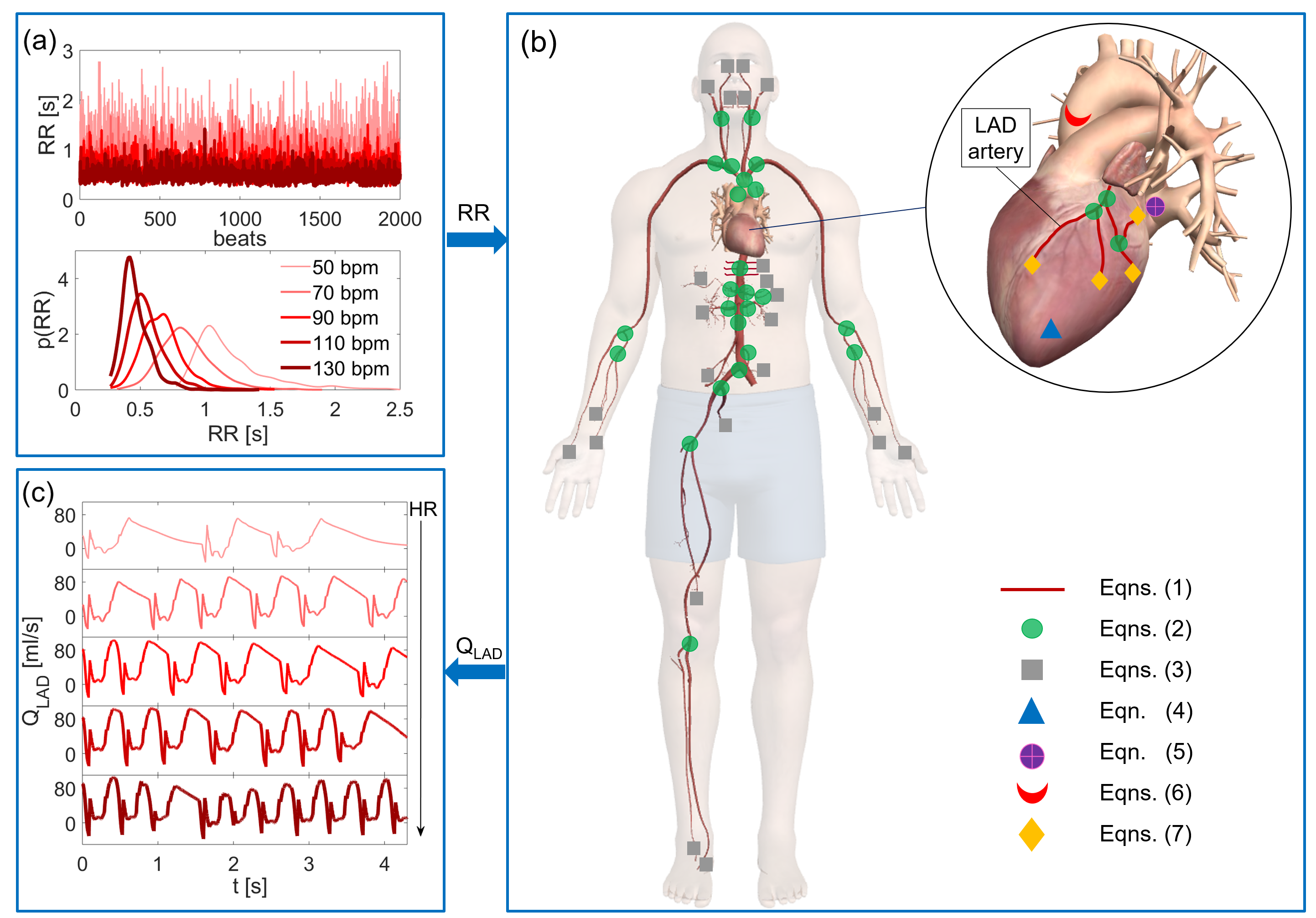}
	\caption{Flow chart of the computational algorithm. (a) RR beating. 2000 RR beats extracted in AF at the selected HR (50, 70, 90, 110, 130 bpm) and the corresponding probability distribution functions. (b) Schematic representation of the 0D-1D cardiovascular model, together with the corresponding governing equations of each district. (c) Examples of the resulting $Q_{LAD}$ time-series at different HRs. \label{flow_chart}}
\end{figure}

\subsection{RR beating features}

RR [s] is defined as the cardiac beating period, with the heart rate, HR=60/RR, expressed in [bpm]. We use artificially built RR intervals to avoid the patient-specific details (e.g., sex, age, weight, and cardiovascular diseases) inherited by real RR beating. RR intervals during AF are described by the superposition of two indipendent times, $RR=\varphi + \eta$. $\varphi$ is taken from a Gaussian distribution and the extraction relies on the correlated pink noise. $\eta$ is taken from an Exponential distribution (with rate parameter $\gamma$) and the extraction is based on the uncorrelated white noise. The resulting RR is thus represented by an exponentially modified Gaussian distribution, which is the most common distribution during AF \cite{Hennig2006,Hayano1997,Scarsoglio2014}. By varying the HR from 50 to 130, each distribution is built keeping the coefficient of variation, $cv=\sigma/\mu$ ($\sigma$ is the standard deviation, $\mu$ is the mean value of the RR distribution), constant and equal to 0.24, which is typically observed during AF \cite{Tateno2001}. The rate parameter, $\gamma$, is approximated as a linear function of the mean RR ($\gamma=-9.2 RR + 14.6$), as estimated from 84 long-term ECG recordings of paroxysmal AF \cite{Anselmino2017}. The in-silico RR intervals have been tested and validated over clinically measured data \cite{Hennig2006,Hayano1997,Sosnowski2011,Goldberger2000}, thus we adopt them as the most suitable and reliable RR time-series to mimic AF conditions. Mean, standard deviation and rate parameter values of all configurations here considered (from 50 to 130 bpm) are reported in Table 1, while probability distributions and time-series of the extracted RR are shown in Fig. 1a.

\begin{table}
\begin{center}
\begin{tabular}{|c|c|c|c|c|c|}
  \hline
   & 50 bpm & 70 bpm & 90 bpm & 110 bpm & 130 bpm \\
   \hline
  $\mu$ [s] & 1.20 & 0.86 & 0.67 & 0.55 & 0.46 \\
  \hline
  $\sigma$ [s]  & 0.29 & 0.21 & 0.16 & 0.13 & 0.11 \\
  \hline
  $\gamma$ [Hz]  & 3.56 & 6.71 & 8.47 & 9.58 & 10.35 \\
  \hline
\end{tabular}
\caption{Mean ($\mu$), standard deviation ($\sigma$), and rate parameter ($\gamma$) values of the RR distributions at the different HRs considered (the resulting cv values are constant and equal to 0.24).}
\end{center}
\end{table}

\noindent For each HR simulation, the multiscale modelling is integrated in time until the main statistics (mean and standard deviation) of the hemodynamic variables are insensitive to further temporal extension; 2000 cardiac periods guarantee the statistical reliability of the results to be reached. Thus, all the hemodynamic variables presented in the Results, as well as the reference heart rate of each configuration (50, 70, 90, 110, 130 bpm), are intended as mean values averaged over 2000 cycles.

\subsection{Mathematical modelling}

The present study relies on the multiscale model proposed by Guala et al. \cite{Guala2017}, which combines the 0D-1D modelling of the left heart and arterial tree \cite{Guala2014,Guala2015,Scarsoglio2018} together with the coronary circulation model as introduced by Mynard et al. \cite{Mynard2014,Mynard2015}. The geometrical domain includes the arterial tree, the left heart, with mitral and aortic valves, and the left coronary circulation. Figure 1b shows a schematic representation of the cardiovascular model, together with the corresponding governing equations of each district. A brief description of the heart-arterial-coronary modelling is recalled hereunder, while more details are offered elsewhere \cite{Guala2014,Guala2017,Mynard2014,Mynard2015}.

\noindent \textbf{Arterial tree}

The systemic arterial circulation consists of 48 large-medium arteries and 24 distal groups. Both left and right arm vessels are taken into account because of their asymmetric geometry, while only the right leg is introduced exploiting the symmetry in leg arteries. The large-medium arteries are described through the 1D form of the mass and momentum equations, in terms of pressure, section area, and flow rate. For each 1D artery, a constitutive equation relating pressure and area completes the differential system

	\begin{equation}
	\begin{dcases}
	\frac{\partial A}{\partial t}+\frac{\partial Q}{\partial x} = 0,  \\
	\frac{\partial Q}{\partial t}+\frac{\partial}{\partial x}\left(\beta\frac{Q^2}{A}\right) = -\frac{A}{\rho}\frac{\partial P}{\partial x}+N_{4}, \label{1D} \\
    P = B_1 + B_2 A + B_3 A^2 + B_4 A^3 - B_5 \frac{1}{\sqrt{A}} \frac{\partial Q}{\partial x}, \\
	\end{dcases}
	\end{equation}

\noindent where $x$ is the spatial coordinate along the vessel axis, $t$ is time, $A$ is the vascular section, $Q$ is the flow rate, $P$ is the radially-constant pressure, $\rho$ is the blood density, $\beta$ is the Coriolis coefficient, $N_{4}$ is the viscous term, and coefficients $B_i$ (i = 1,...,5) contain the information about the local viscoelastic mechanical properties.

Conservation of mass and total pressure is set at each arterial junction (the subscript 0 refers to the parent artery, while the subscripts 1 and 2 are associated to the daughter vessels)
	
	\begin{equation}
	\begin{dcases}
			Q_{0} = Q_{1}+Q_{2}, \\
			P_{0}+\frac{1}{2}\rho\bar{u}_{0}^2 = P_{1}+\frac{1}{2}\rho\bar{u}_{1}^2, \label{bifurcation} \\
			P_{0}+\frac{1}{2}\rho\bar{u}_{0}^2 = P_{2}+\frac{1}{2}\rho\bar{u}_{2}^2, \\
	\end{dcases}
	\end{equation}

\noindent where $\bar{u}$ is the mean blood velocity.

The distal groups consist of three-element Windkessel submodels

	\begin{equation}
	\frac{\partial 	Q_{b}}{\partial t}-\frac{1}{R_{1}}\frac{\partial P_{b}}{\partial t}=\frac{P_{b}-P_{ven}}{R_{1}R_{2}C}-\left(1+\frac{R_{1}}{R_{2}}\right)\frac{Q_{b}}{R_{1}C},
	\label{WK3}
	\end{equation}
	
\noindent where $Q_{b}$ and $P_{b}$ are the flow rate and pressure at the outlet of the generic terminal artery, $R_{1}$ equals the characteristic impedance of the artery, $R_{2}$ and $C$ represent the viscous and elastic features of the distal area, respectively. These groups form the outlet boundary conditions, accounting for the missing cerebral circulation at the terminal carotid and vertebral arteries, the peripheral micro-circulation, the venous return and the right heart.

\noindent \textbf{Left heart}

Left ventricle is described through a time-varying elastance model,

	\begin{equation}
	E_{LV}=\frac{P_{LV}}{V_{LV}-V_{0}} \label{Elv},
	\end{equation}

\noindent which relates left ventricular pressure, $P_{LV}$, and volume, $V_{LV}$ ($V_{0}$ is the unstressed left ventricular volume). Mitral valve is simulated as an ideal diode with resistance $R_{MI}$, and the mitral flow, $Q_{MI}$, is ruled by the pressure gradient across the valve ($P_{LA}$ is the left atrial pressure):

	\begin{equation}
	\begin{dcases}
	Q_{MI}=\frac{P_{LA}-P_{LV}}{R_{MI}},  \,\,\, \textmd{if} \,\,\, P_{LA}>P_{LV}, \label{mi} \\
	Q_{MI}=0,  \,\,\, \textmd{if} \,\,\, P_{LA} \leq P_{LV}.\\
	\end{dcases}
	\end{equation}

\noindent The aortic valve is instead modeled by means of a pressure-flow rate relation which, through the valve opening angle $\theta(t)$, accurately accounts for blood flow effects such as forces due to the pressure difference across the valve ($F_{pr}$), frictional effects from neighboring tissue resistance ($F_{fr}$), the dynamic effect of the blood acting on the valve leaflet ($F_{bm}$), and the action of the vortex downstream of the valve ($F_{vo}$) \cite{Korakianitis2006}:

	\begin{equation}
	\begin{dcases}
	L\frac{dQ_{AA}}{dt}+RQ_{AA}+B|Q_{AA}|Q_{AA} = \Theta\left(P_{LV}-P_{AA}\right),  \\
	\Theta = \frac{\left(1-\cos(\theta)\right)^4}{\left(1-\cos(\theta_{max})\right)^4}, \label{ao} \\
	I_{ao}\frac{d^2\theta}{dt^2} = F_{pr} - F_{fr} + F_{bm} - F_{vo}, \\
	\end{dcases}
	\end{equation}

\noindent where $Q_{AA}$ and $P_{AA}$ are the flow rate and pressure at the entrance to the aorta, $L$, $R$ and $B$ stand for the inertance, viscous and turbulent flow separation effects, respectively. The function $\Theta$ simulates the nonideal behaviour of the valve orifice, while $\theta_{max}$ is the maximum aortic valve opening angle (here, $\theta_{max}$=75$^{\circ}$), and $I_{ao}$ is the rotational inertia of the aortic leaflets.

\noindent The adoption of the pressure-flow rate relation (\ref{ao}) instead of an aortic diode model allows us to include important arterial flow features, such as the physiologic back flow in the aortic valve following its closure and the occurrence of the dicrotic notch, which would not be captured by the simplified model. In fact, this latter, by only considering unidimensional flow, is not able to reproduce the complexity of the valvular flow, such as the presence of stenotic or regurgitant flow, and sinus vortices downstream of the valve \cite{Korakianitis2006,Aboelkassem2015}. Moreover, the diode model does not account for geometry variations \cite{Aboelkassem2015}, as well as accurate pressure effects with respect to the current state (or position) of the valve \cite{Korakianitis2006,Mynard2012}.

\noindent Equations (\ref{Elv}), (\ref{mi}), and (\ref{ao}) represent the complete set of inlet boundary conditions of the arterial tree.

\noindent \textbf{Left coronary circulation}

The left coronary circulation originates from the aortic root and includes 7 large-medium arteries, which are modelled as the systemic arterial tree (Eqns \ref{1D}). Each artery ends up with a microvascular distal district, identified at the entrance by a characteristic impedance, $Z_{pa}$, pressure, $P_a$, and compliance, $C_{pa}$, of the corresponding 1D coronary artery, and composed by three transmural layers: the subepicardium (j=1), the midwall (j=2), and the subendocardium (j=3). Each layer is forced by an intramyocardial pressure, $P_{j}^{im}$, and is divided into two compartments, the arterial one and the venous one. Each compartment is described by a resistance and a compliance, $R_{1j}$ and $C_{1j}$ are the arterial resistances and compliances, while $R_{3j}$ and $C_{2j}$ are the venous resistance and compliances. An intermediate resistance, $R_{2j}$, matches the arterial and venous compartments, with $P_{1j} - P_{2j}$ being the pressure difference through it. Compliances are constant in time, while all resistances vary in time, based on instantaneous volume filling of each sublayer. The time-varying resistance mimics the autoregulation capability to supply the appropriate blood flow level to the heart in physiological conditions. The venous compartments of the microvascular model converge into a venula, which is characterized - as the entrance - by a specific impedance, $Z_{pv}$, pressure, $P_v$, and compliance, $C_{pv}$. Pressure at the outlet of the coronary microcirculation model, $P_{out}$, is taken constant and equal to 5 mmHg. Combining constitutive and conservation equations, each coronary distal group is ruled by the following system

	\begin{equation}
	\begin{dcases}
		C_{1j}\frac{d P_{1j}}{dt} = C_{1j}\frac{dP_{j}^{im}}{dt}+\frac{P_{a}-P_{1j}}{R_{1j}}-\frac{P_{1j}-P_{2j}}{R_{2j}}, \,\,\,\,  j=1,...,3, \\
		C_{2j}\frac{d P_{2j}}{dt} = C_{2j}\frac{dP_{j}^{im}}{dt}+\frac{P_{1j}-P_{2j}}{R_{2j}}-\frac{P_{2j}-P_{v}}{R_{3j}}, \,\,\,\,  j=1,...,3, \\
		C_{pa}\frac{d P_a}{dt} = \frac{P_{in}-P_{a}}{Z_{pa}}-\sum_{j=1}^3 \frac{P_{a}-P_{1j}}{R_{1j}}, \label{distal_coronary} \\
		C_{pv}\frac{d P_v}{dt} = \sum_{j=1}^3 \frac{P_{2j}-P_{v}}{R_{3j}}-\frac{P_{v}-P_{out}}{Z_{pv}}, \\	
	\end{dcases}
	\end{equation}

\noindent where all parameters are set as in \cite{Mynard2014,Mynard2015}.

The complete heart-arterial-coronary modelling results in a system of hyperbolic partial differential equations, which is numerically solved through a Discontinuous-Galerkin Runge-Kutta scheme. Parameter setting relative to the coupling of different models relies to a great extent on what proposed by \cite{Guala2017}. Further tuning of the geometrical and mechanical parameters was guided by achieving reliable coronary flow rate signals in normal sinus rhythm, as obtained by \cite{Mynard2014,Mynard2015}.

\subsection{Definition of hemodynamic parameters for coronary perfusion}

We here introduce the hemodynamic parameters used to quantify coronary blood flow variations at different HRs during AF. In particular, all parameters are computed based on the left anterior descending (LAD) artery, which is the most important vessel in terms of myocardial blood supply (see Fig. 1c). By dividing the RR interval into systolic, $RR_{sys}$, and diastolic, $RR_{dia}$, periods, we can define $V_{sys}$ [ml/beat] as the volume of blood flowing through the LAD artery during the systole (see Fig. 2)

\begin{equation}
V_{sys}=\int_{RR_{sys}} Q_{LAD}(t)dt,
\end{equation}

\noindent where $Q_{LAD}$ is the flow rate signal in the LAD artery. Similarly, $V_{dia}$ [ml/beat] is introduced as the volume of blood flow through the LAD artery during the diastole (see Fig. 2)

\begin{equation}
V_{dia}=\int_{RR_{dia}} Q_{LAD}(t)dt.
\end{equation}

\noindent The combination of $V_{sys}$ and $V_{dia}$ gives the net flow through the LAD artery, defined as LAD stroke volume, $SV = V_{sys} + V_{dia}$ [ml/beat].
By means of SV, the coronary blood flow per beat through the LAD artery is expressed as CBF=SV x HR [ml/min], giving a measure of myocardial  perfusion and oxygen supply. Other  quantities related to the LAD flow rate are $Q_{max,sys}$, defined as the maximum flow rate during $RR_{syst}$, and $Q_{max,dia}$, which is the maximum flow rate in the $RR_{dia}$ interval. $Q_{min}$ is instead defined as the absolute minimum flow rate value reached during $RR$. The hemodynamic parameters introduced so far are all beat-to-beat measures, that is for each $RR$ beat a value is obtained. In the end, a 2000 data distribution is available for each parameter at a fixed HR configuration. A typical LAD flow rate signal, $Q_{LAD}$, is reported for a single AF beat (0.87 [s]) in Fig. 2, together with indication of $V_{sys}$, $V_{dia}$, $Q_{max,dia}$, $Q_{max,sys}$, and $Q_{min}$.

\begin{figure}[h!]
	\centering
	\includegraphics[width=0.9\columnwidth, trim=45 35 85 520, clip=true]{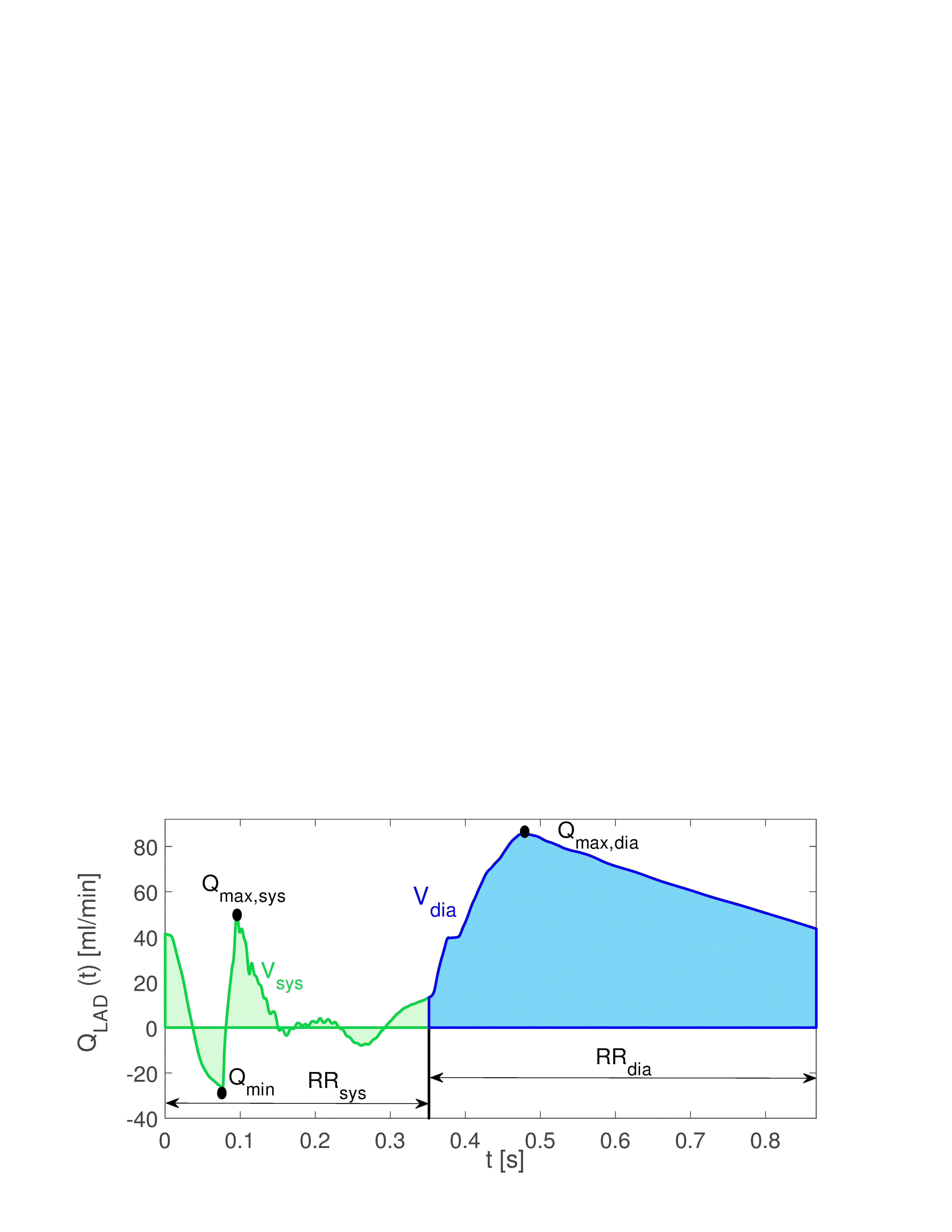}
	\caption{Example of LAD flow rate, $Q_{LAD}$, during AF (RR=0.87 [s]). $Q_{min}$ is the minimum flow rate, $Q_{max,sys}$ and $Q_{max,dia}$ are the maximum flow rates during the systolic $RR_{sys}$ and diastolic $RR_{dia}$ periods, respectively. $V_{sys}$ (green) and $V_{dia}$ (blue) are the blood volumes per beat during $RR_{sys}$ and $RR_{dia}$, respectively. \label{LAD_waveform}}
\end{figure}

To better compare the different HRs, we also define the average behaviour of the LAD flow rate per beat over $N$ cardiac periods

\begin{equation}
<Q_{LAD}>=\frac{1}{N}\sum_{i=1}^{N} Q_{LAD_{i}}(\tau),
\label{Q_{LAD}}
\end{equation}

\noindent where here $N=2000$, $Q_{LAD_{i}}$ is the $Q_{LAD}$ of the i-th heartbeat ($i=1,...,N$), while $\tau=t/RR_i$ is the non-dimensional temporal coordinate ($\tau \in [0, 1]$). For each HR configuration, $<Q_{LAD}>$ is introduced to highlight the average LAD flow rate waveform over 2000 cardiac cycles and expressed in terms of the non-dimensional beating period, $\tau$. Beside $<Q_{LAD}>$, we can also evaluate the standard deviation of the $Q_{LAD_{i}}$ signals, namely $\sigma_{Q_{LAD}}$. Through the adimensional time $\tau$,  both $<Q_{LAD}>$ and $\sigma_{Q_LAD}$ vary in the temporal interval [0, 1] for all the HRs, and are therefore fully comparable.

We also introduce the coronary perfusion pressure, $CPP=P_{dia} - P_{lved}$, where $P_{dia}$ is the diastolic aortic pressure, while $P_{lved}$ is the end-diastolic left-ventricular pressure. CPP is usually referred to as a surrogate measure for myocardial perfusion \cite{BerneLevy}. In the end, to evaluate the myocardial oxygen supply-demand ratio, we compute the rate pressure product, RPP = P$_{sys}$ x HR [mmHg/min] (where $P_{sys}$ is the systolic arterial pressure), which is an estimate of the oxygen consumption \cite{Westerhof}.

\section{Results}

\subsection{Waveform and amplitude of the coronary blood flow rate}

We first focus on how waveform and amplitude of the coronary blood flow rate change with HR during AF. Fig. 3 shows the different $<Q_{LAD}>$ flow rates, together with the corresponding $\sigma_{Q_{LAD}}$, while Table 2 reports $Q_{max,dia}$,  $Q_{max,sys}$, and $Q_{min}$ at different HRs. It should be recalled that Fig. 3 and Table 2 highlight different aspects of the signal shape: the average LAD flow rate represents the characteristic mean waveform per beat as averaged over 2000 beats, while Table 2 reports the basic statistics of minima and maxima distribution, as measured over 2000 beats. For this reason, the average LAD flow rate definition is mainly exploited to compare waveform variations at different HRs, while minima and maxima of the flow rate signal are monitored to quantify amplitude variations.

\begin{figure}[h!]
\begin{minipage}[]{0.5\columnwidth}
\includegraphics[width=\columnwidth, trim=0 0 210 495, clip=true]{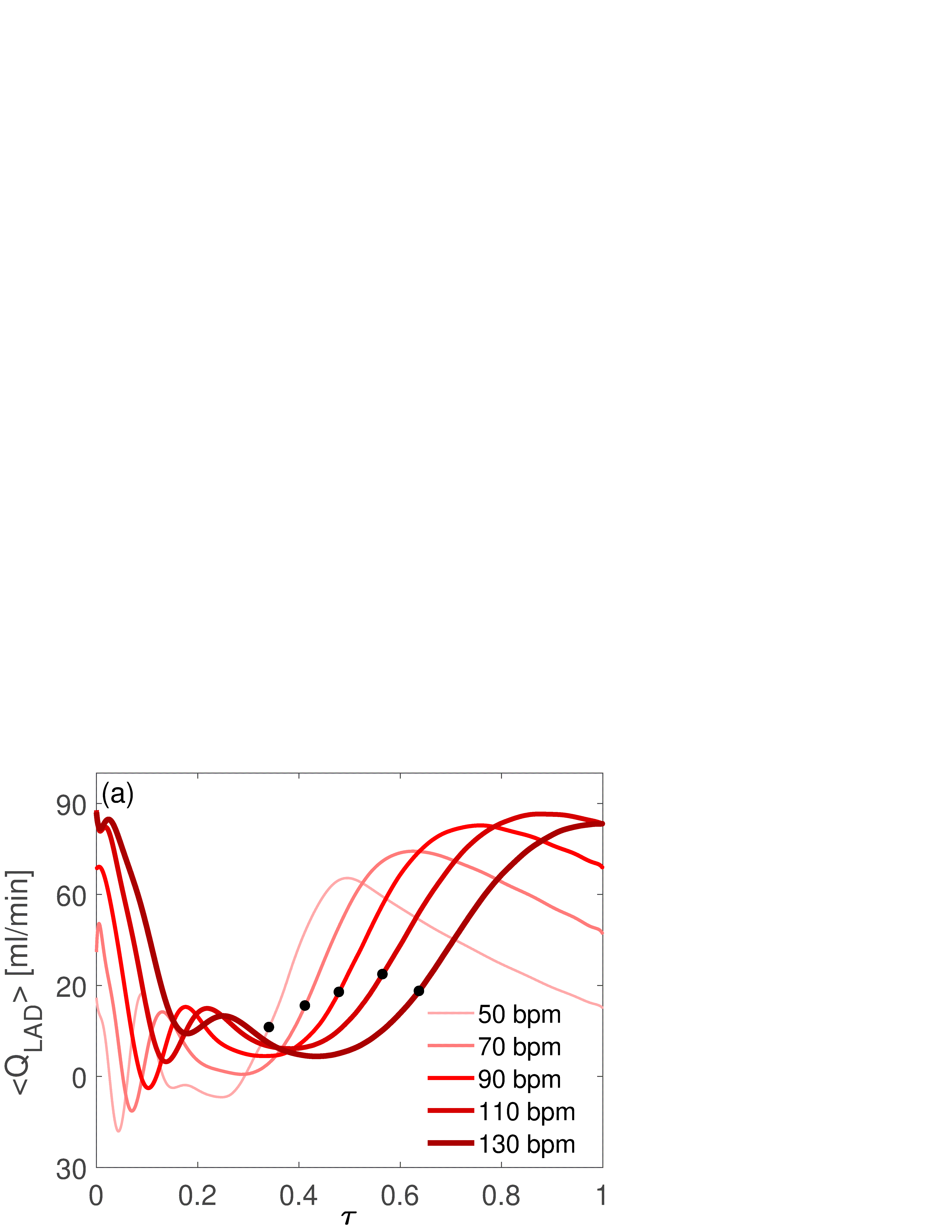}
\end{minipage}
\begin{minipage}[]{0.5\columnwidth}
\includegraphics[width=\columnwidth, trim=0 0 210 495, clip=true]{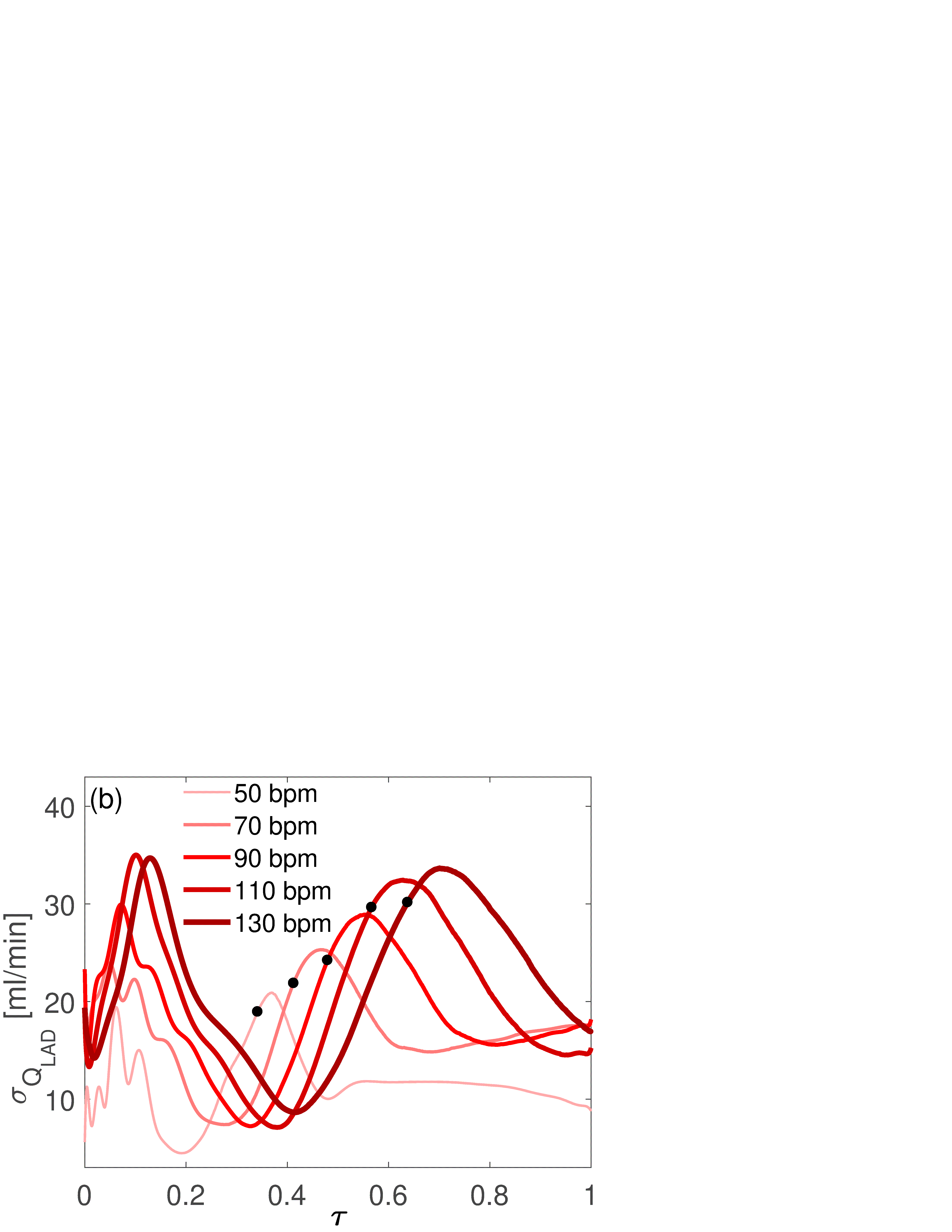}
\end{minipage}
\caption{(a) Average ($<Q_{LAD}>$) and (b) standard deviation ($\sigma_{Q_{LAD}}$) LAD flow rate signals at different HRs (50, 70, 90, 110, 130 bpm) as function of the non-dimensional time, $\tau$. Black dots indicate the beginning of the diastole.}
\end{figure}

Starting with the waveform, Fig. 3a shows that by increasing HR the average waveform is deeply modified. During both systole and diastole, at higher HRs the $<Q_{LAD}>$ signal is remarkably stretched forward in time with respect to lower HRs. As for the diastole, the typical behaviour revealed at lower HRs (also shown in Fig. 1c and 2) with a rapid growth, a maximum peak and a slow decrease, is completely lost at higher HRs. This can be explained considering that, at higher HR, the diastole length is much more reduced with respect to systole, with the ratio $RR_{dia}/RR_{sys}$ ranging from 1.93 at 50 bpm to 0.57 at 130 bpm. Therefore, for HR $>$ 90 bpm $<Q_{LAD}>$, which is ruled by the diastolic phase, strongly changes its typical features: the diastolic growth of $<Q_{LAD}>$ is much slower, so that the maximum peak is reached towards the end of the beat, without even facing the decreasing phase. As a consequence, at higher HRs the next systolic phase starts with much higher $<Q_{LAD}>$ values.

\noindent The $\sigma_{Q_{LAD}}$ curves at different HRs are also reported in Fig. 3b. $\sigma_{Q_{LAD}}$ overall grows with HR and presents a time lag for increasing HR similar to $<Q_{LAD}>$. Moreover, there is a maximum of $\sigma_{Q_{LAD}}$ for each of the two cardiac phases: these maxima occur where significant beat-to-beat variations emerge. With reference to $Q_{min}$, $Q_{max,sys}$, and $Q_{max,dia}$ (as displayed in Fig. 2), the systolic maximum of $\sigma_{Q_{LAD}}$ indicates the temporal range of rapid variations between $Q_{min}$ and $Q_{max,sys}$, while the diastolic maximum corresponds to the temporal region of rapid growth before reaching $Q_{max,dia}$.

\begin{table}[h!]
	\centering
	\begin{tabular}{|c|c|c|c|} \hline
		HR [bpm] & $Q_{max,sys}$ [ml/min] & $Q_{max,dia}$ [ml/min] & $Q_{min}$ [ml/min]\\
        \hline
		$50$ & $48.97\pm5.36$ & $73.92\pm7.07$ & $-27.65\pm3.54$ \\
		     &($0.11$)        &($0.096$)       & $(0.13)$ \\
        \hline
		$70$ & $56.36\pm10.65$ & $86.75\pm8.60$ & $-28.06\pm3.03$\\
		     &($0.19$)         & ($0.099$)      & $(0.11)$ \\
        \hline
		$90$ & $71.11\pm15.07$ & $94.30\pm9.70$ & $-27.85\pm4.02$\\
		     &($0.21$)         & ($0.10$)       & $(0.14)$ \\
        \hline
		$110$ & $84.68\pm15.08$ & $96.61\pm11.15$ & $-27.49\pm3.81$\\
		      &($0.18$)         & ($0.12$)        & $(0.14)$ \\
        \hline
		$130$ & $87.41\pm14.85$ & $89.52\pm17.24$ & $-25.91\pm5.45$\\
		      & ($0.17$)        & ($0.19$)        & $(0.21)$ \\
        \hline
	\end{tabular}
	\caption{Mean ($\mu$) and standard deviation ($\sigma$) values of $Q_{max,sys}$, $Q_{max,dia}$, and $Q_{min}$ over 2000 beats for each simulated cardiac frequency (50, 70, 90, 110, 130 bpm) in AF. Coefficients of variation ($cv=\sigma/\mu$) are indicated in brackets. \label{max_min_flow_rates}}
\end{table}

Considering the amplitude variation of the LAD flow rate signal, $Q_{LAD}$, Table 2 reports mean, standard deviation and coefficient of variation values for the distribution of $Q_{max,dia}$,  $Q_{max,sys}$, and $Q_{min}$, over 2000 cardiac beats at different HRs.
Mean values of the maxima ($Q_{max,sys}$ and $Q_{max,dia}$) grow with HR: the highest and monotone increase is found for $Q_{max,sys}$ (+78\% from 50 to 130 bpm), while $Q_{max,dia}$ increases by 21\% (from 50 to 130 bpm), presenting a maximum value around 110 bpm. The growth of $Q_{min}$ in average is instead much less pronounced, i.e. +6\% (from 50 to 130 bpm). cv values of the three parameters overall increase ranging from 50 to 130 bpm, turning out to be quite sensitive to HR variability, especially recalling that cv of the RR series is maintained constant at all HR configurations.

\subsection{Coronary perfusion}

\begin{figure}[h!]
\begin{minipage}[]{0.5\columnwidth}
\includegraphics[width=\columnwidth, trim=0 0 185 475, clip=true]{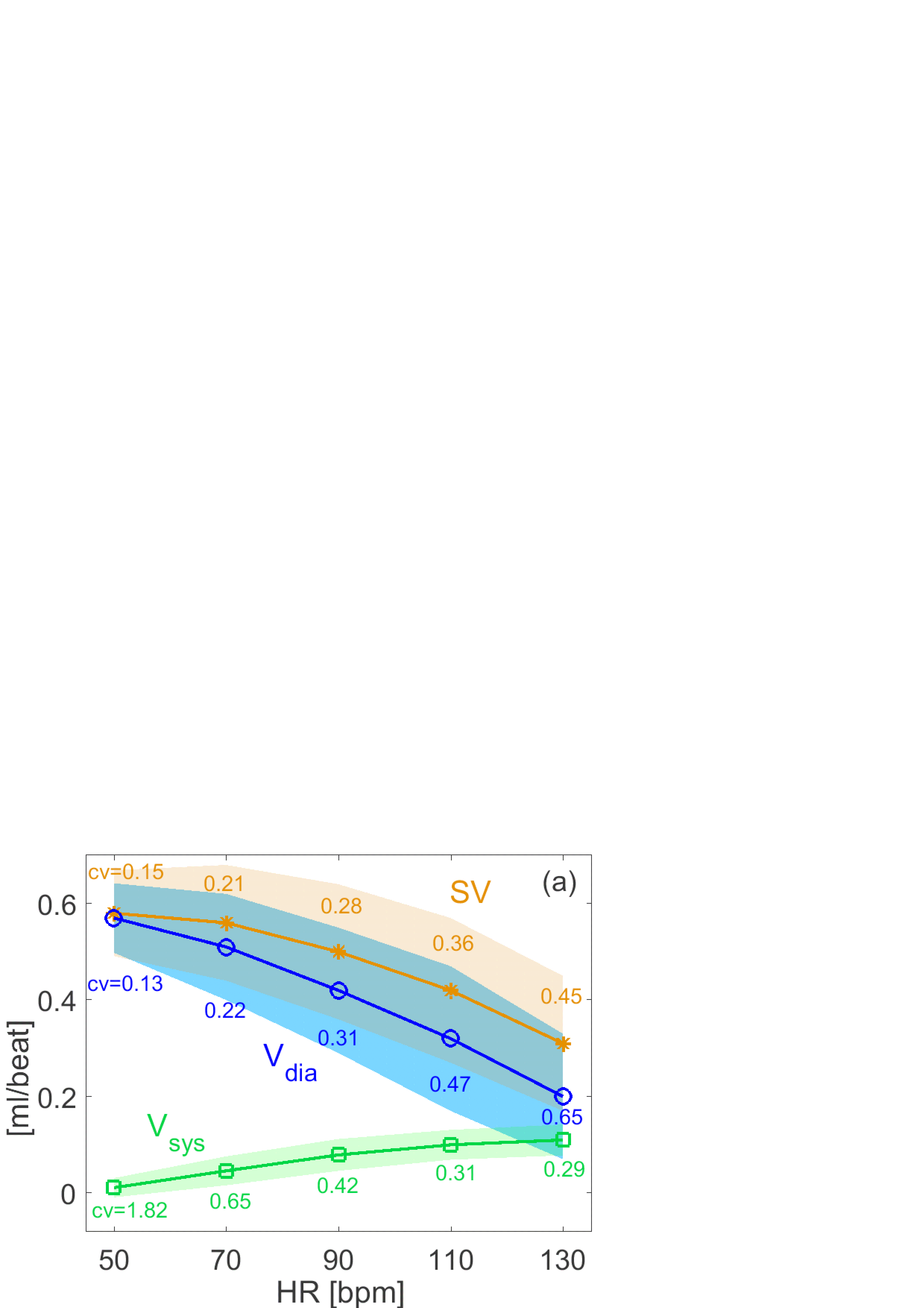}
\end{minipage}
\begin{minipage}[]{0.5\columnwidth}
\includegraphics[width=\columnwidth, trim=0 0 185 475, clip=true]{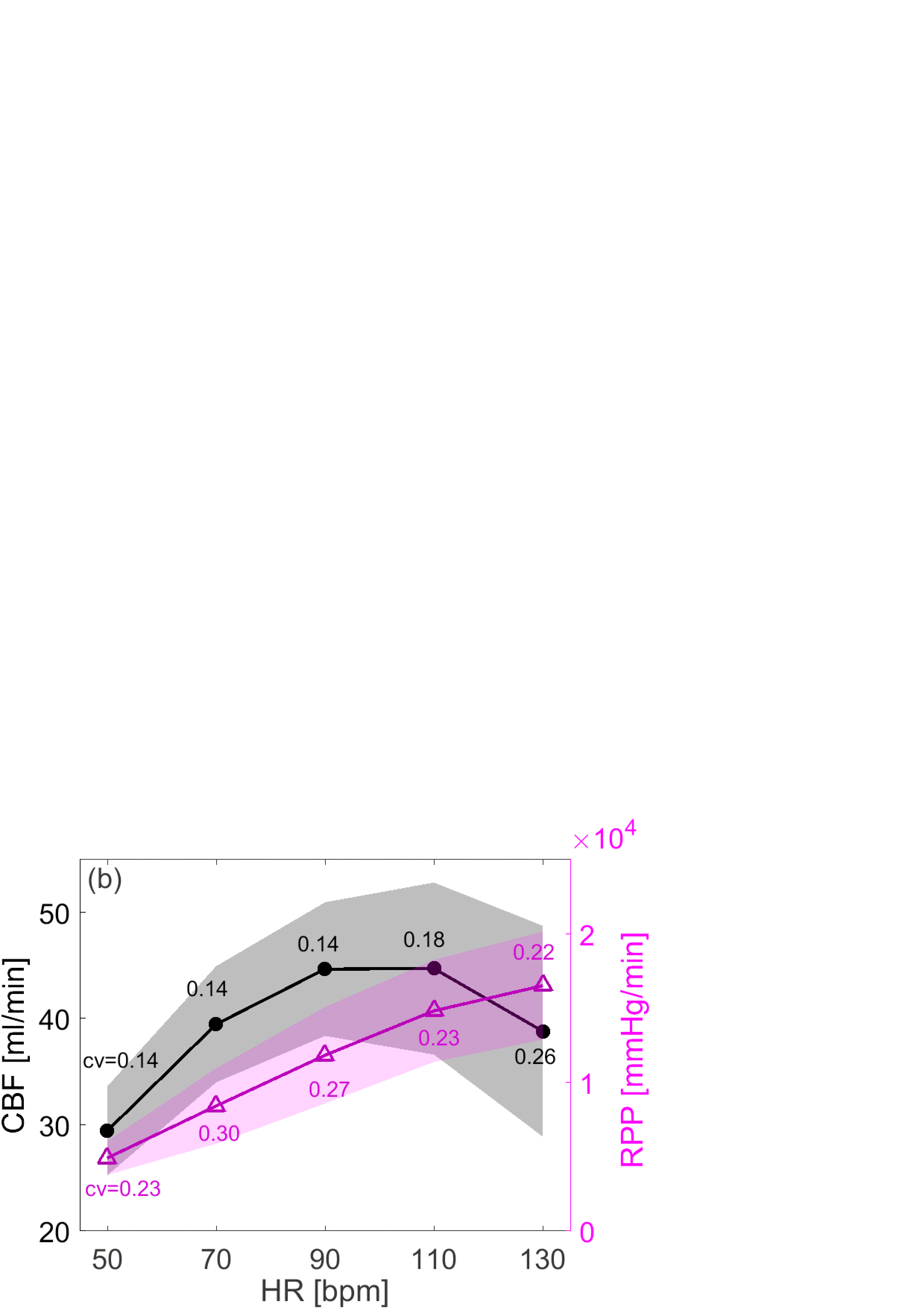}
\end{minipage}
\caption{Mean values ($\mu$, solid curves) and standard deviation values (in terms of $\mu\pm\sigma$, shaded areas) of hemodynamic parameters as function of HR (computed over 2000 beats): (a) $V_{sys}$, $V_{dia}$, $SV$; (b) CBF (left y-axis), RPP (right y-axis). $cv=\mu/\sigma$ values are also reported for each variable and HR.}
\end{figure}

The coronary perfusion is analyzed in terms of the coronary blood flow, $CBF$, and the hemodynamic variables related to it. In Fig. 4 mean values ($\mu$) and standard deviation values (in terms of $\mu \pm \sigma$) for $V_{sys}$, $V_{dia}$, $SV$, $CBF$, and $RPP$ at different HRs are reported with solid curves and shaded areas, respectively. For increasing $HR$, a sharp $V_{dia}$ decrease and a slight $V_{sys}$ increase in absolute terms occur, thus $SV$ overall decreases with $HR$. However, the diastolic volume decrease does not proportionally scales with HR, but affects mostly higher HRs. This results in a non-monotonic behaviour of CBF=SV x HR, which shows a maximum around 90-110 bpm and then a decrease for higher HR. The oxygen demand, expressed by $RPP$, monotonically grows with HR.

\noindent Fig. 4 also displays cv values which, apart from $V_{sys}$ and $RPP$, all increase with HR. Beside the average trend just described for SV and CBF, it is important to remark that for these two hemodynamic parameters the AF-induced variability monotonically increases with HR, despite the input RR variability is imposed as constant for all the HRs.

\begin{figure}[h!]
\begin{minipage}[]{0.329\columnwidth}
\hspace{-.2cm}
\includegraphics[width=\columnwidth, trim=0 0 210 480, clip=true]{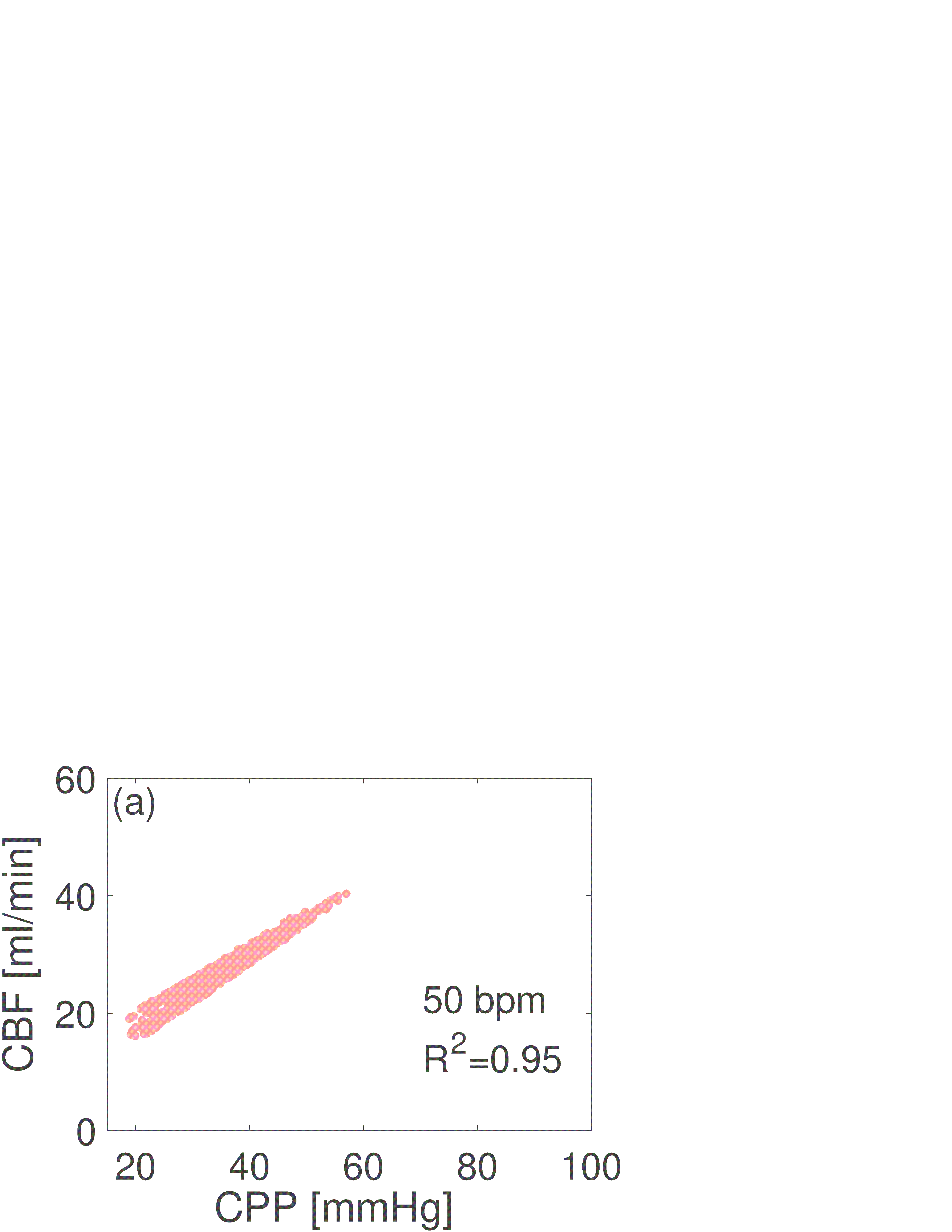}
\end{minipage}
\begin{minipage}[]{0.329\columnwidth}
\hspace{-.2cm}
\includegraphics[width=\columnwidth, trim=0 0 210 480, clip=true]{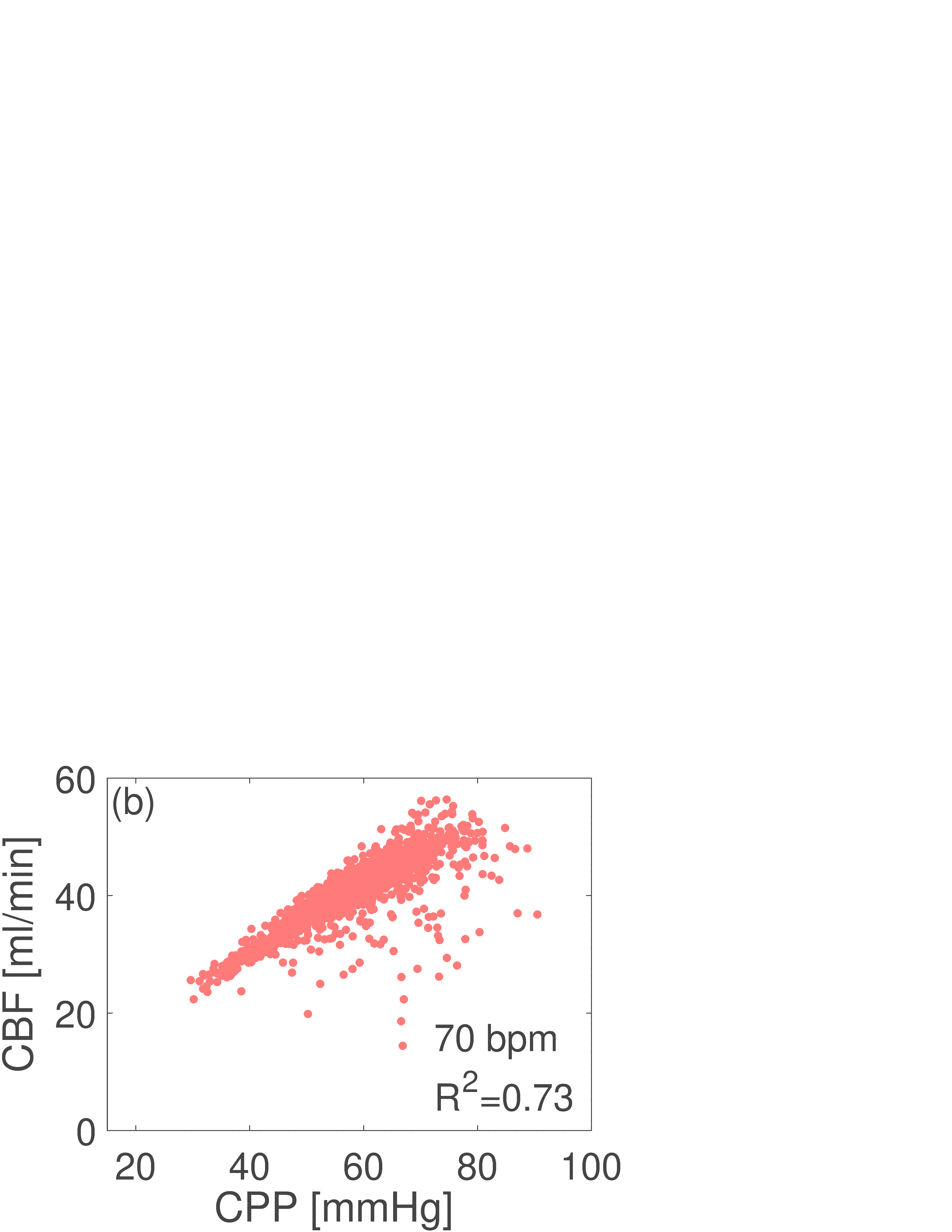}
\end{minipage}
\begin{minipage}[]{0.329\columnwidth}
\hspace{-.2cm}
\includegraphics[width=\columnwidth, trim=0 0 210 480, clip=true]{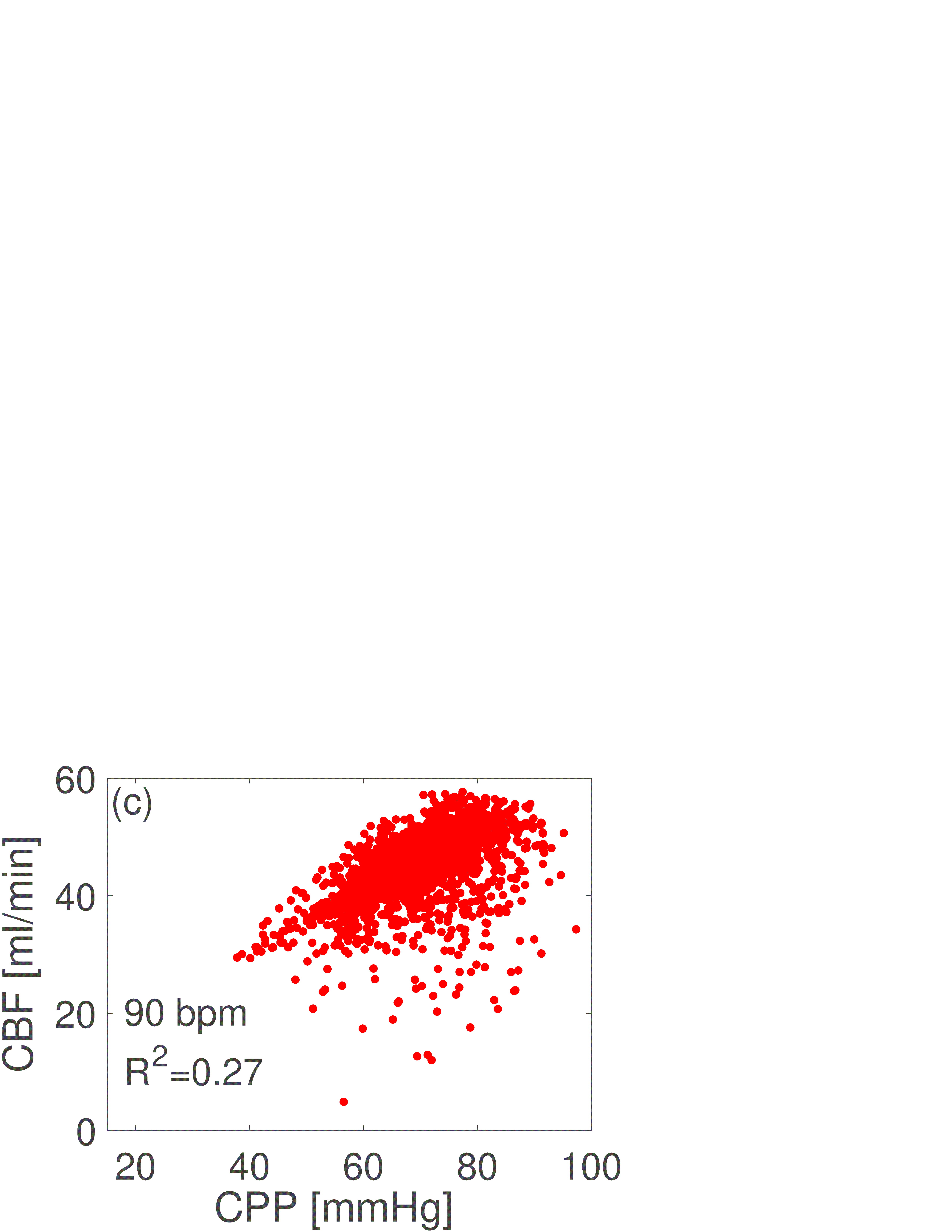}
\end{minipage}
\begin{minipage}[]{0.329\columnwidth}
\hspace{-.2cm}
\includegraphics[width=\columnwidth, trim=0 0 210 480, clip=true]{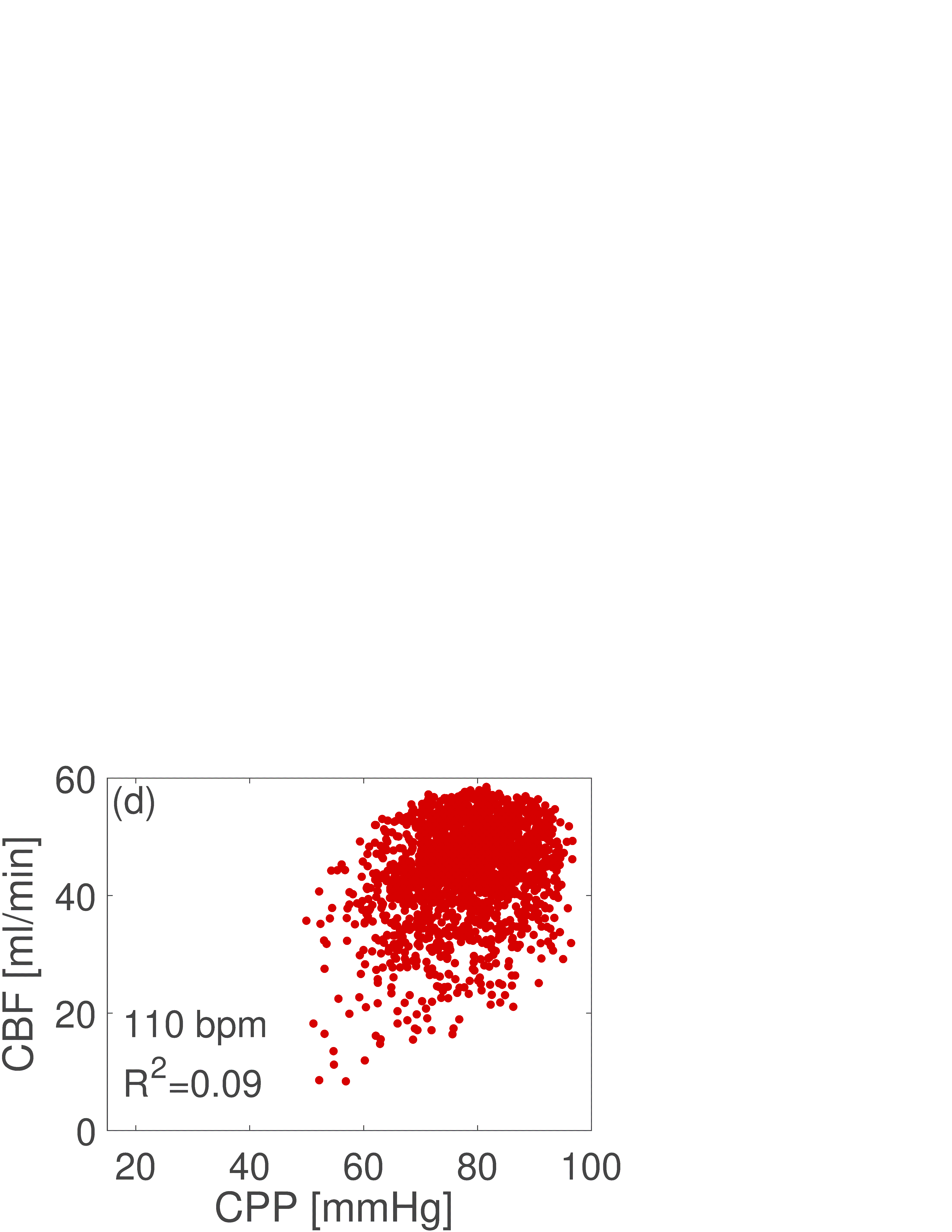}
\end{minipage}
\begin{minipage}[]{0.329\columnwidth}
\hspace{-.2cm}
\includegraphics[width=\columnwidth, trim=0 0 210 480, clip=true]{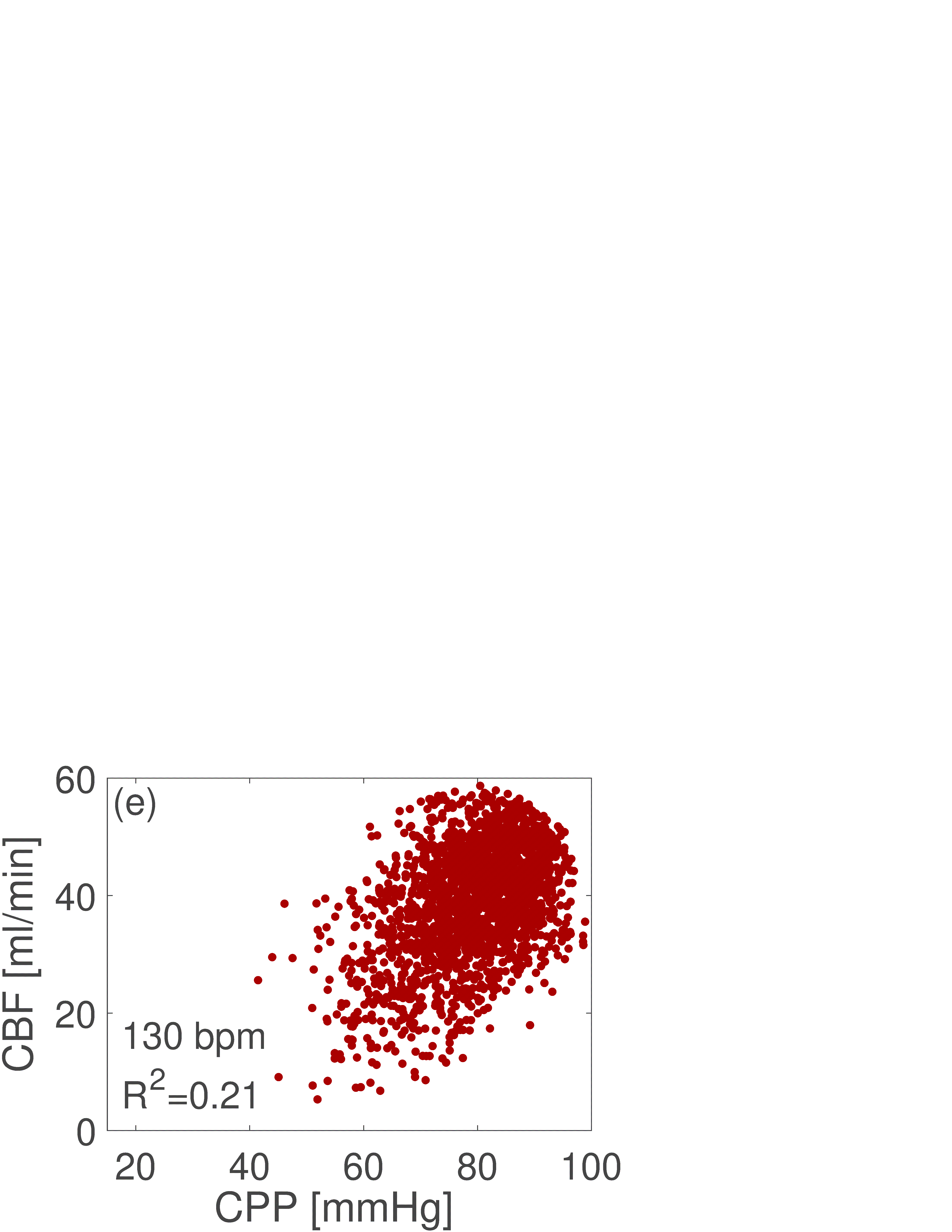}
\end{minipage}
\begin{minipage}[]{0.329\columnwidth}
\hspace{-.2cm}
\includegraphics[width=\columnwidth, trim=0 0 210 480, clip=true]{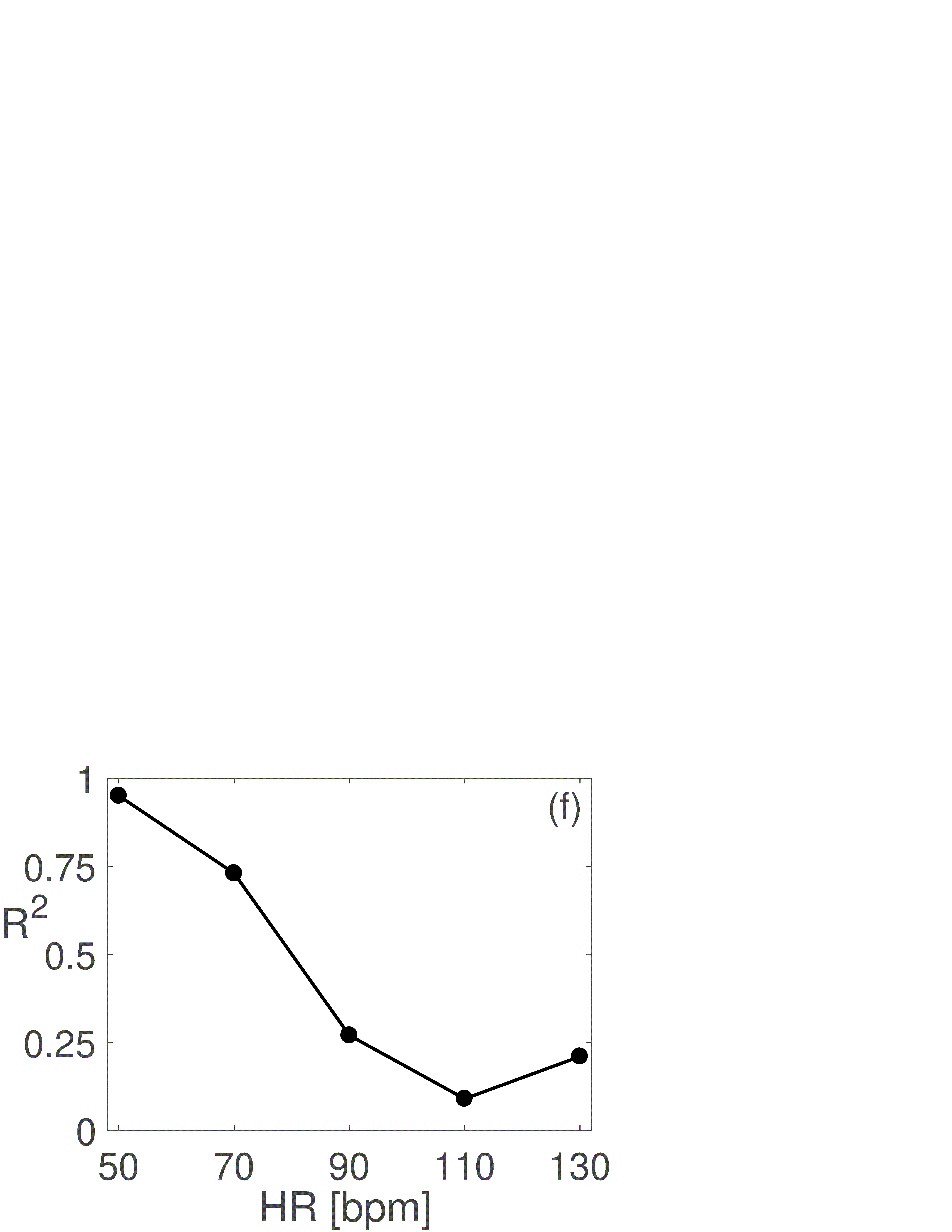}
\end{minipage}
	\caption{Coronary blood flow (CBF) as function of the coronary perfusion pressure (CPP) over 2000 cardiac cycles for HR=50, 70, 90, 110, 130 bpm (panels from a to e). Panel f shows the coefficient of determination, $R^2$, for the relation CBF(CPP) as function of HR.}
\end{figure}

Alongside the basic statistics of the hemodynamics parameters, to further inquire into the irregularity induced by AF, we ask whether coronary perfusion is differently affected by AF variability for increasing HR. To this end, dispersion plots of CBF as function of $CPP=P_{dia} - P_{lved}$ are proposed in Fig. 5, at different HRs (panels from a to e). In normal sinus rhythm, the relation $CBF(CPP)$ is usually exploited to estimate CBF starting from CPP, since they are positively correlated as a result of the coronary autoregulation \cite{BerneLevy}. In AF, we observe that a strong linear correlation between CBF and CPP is still present at lower HR (Fig. 5, panels a and b), while it drops at higher HR (panels c to e). It should be pointed out that, as we assume a constant left atrial pressure, fluctuations of the end-diastolic left-ventricular pressure, $P_{lved}$, are underestimated at any HR. Fig. 5f displays the coefficient of determination, $R^2$, as function of the HR, revealing that for HR$\geq$90 bpm CPP is no longer a good index to approximate the coronary blood flow. In this HR limit, due to the wide data dispersion, a CPP value of 70 mmHg can correspond to CPP ranging from 10 to 55 ml/s.

In the end, to quantify the beat-to-beat variability in CBF during AF, we evaluate how CBF is correlated with the current ($RR$), the preceding ($RR_{-1}$), and the pre-preceding ($RR_{-2}$) heart beats. Table 3 reports the Pearson correlation coefficients, $r$, of each relation ($CBF(RR)$, $CBF(RR_{-1})$, $CBF(RR_{-2})$), stratified over HR. For all the three relations, we move from an inverse to a direct correlation by increasing HR. Moreover, by setting $|r|>0.3$ as significant correlation, CBF results in being mainly correlated with $RR$ and the strength of the correlation decreases considering the preceding ($RR_{-1}$) and the pre-preceding ($RR_{-2}$) beat. In fact, $CBF(RR)$ presents significant correlation at all HRs but 90 bpm, while $CBF(RR_{-2})$ displays a significant correlation only for 70 bpm. This trend is overall in agreement with Kochiadakis et al. \cite{Kochiadakis2012}, who found significant correlation between coronary flow and current $RR$ for all AF patients considered, while scarse and no correlation was found with $RR_{-1}$ and $RR_{-2}$, respectively.

\begin{table}[h!]
\begin{center}
\begin{tabular}{|c|c|c|c|}
  \hline
   HR [bpm] & $CBF(RR)$ & $CBF(RR_{-1})$ & $CBF(RR_{-2})$  \\
   \hline
  50 & -0.70 & -0.66 & -0.29 \\
  \hline
  70 & -0.35 & -0.81 & -0.48 \\
  \hline
  90 & 0.24 & -0.57 &  -0.24 \\
  \hline
  110 & 0.67 & -0.17 & 0.15 \\
  \hline
  130 & 0.73 & 0.05 & 0.28 \\
  \hline
\end{tabular}
\caption{Pearson correlation coefficients, $r$, at different HRs (50, 70, 90, 100, 130 bpm) for the relations $CBF(RR)$ (II column), $CBF(RR_{-1})$ (III column), $CBF(RR_{-2})$ (IV column).}
\end{center}
\end{table}

\section{Discussion}

The coronary impairment at higher HR during AF highlighted by the present findings has a double key to interpretation. We can roughly distinguish coronary alterations mainly imputable to: (i) changes of the heart rate, and (ii) beat-to-beat variability. In details:

\smallskip

(i) we observed that average waveforms, $\langle Q_{LAD}\rangle$, are largely influenced by the HR. In particular, both diastolic and systolic waveform variations are mainly due to HR changes. In fact, as previously mentioned, if HR grows the beat length decreases, with a greater reduction in $RR_{dia}$ than in $RR_{sys}$, ranging the ratio $RR_{dia}/RR_{sys}$ from 1.93 at 50 bpm to 0.57 at 130 bpm. Therefore, at higher frequencies, the typical diastolic decay of $\langle Q_{LAD}\rangle$ observed at smaller HRs disappears, mainly because of a shorter $RR_{dia}$ (see black dots in Fig. 3a indicating the beginning of the diastole at each HR). On the contrary, systole becomes longer than diastole for increasing HR and this results in a wider temporal range over which the high $<Q_{LAD}>$ systolic values can be slowly damped.

\noindent Waveform variations, in turn, influence mean coronary perfusion at different HRs. In fact, the SV reduction with HR is mainly imputable to the diastole reduction as HR grows and is clearly observable from the $<Q_{LAD}>$ waveforms of Fig. 3a: for increasing HR, $Q_{max,dia}$ is quite delayed with respect to the beginning of the diastole and this causes a consistent reduction of the diastolic flow. However, the SV reduction does not proportionally scales with HR, but is slower for lower HR and faster for higher HR. In terms of coronary blood flow (CBF=SVxHR), the slow SV reduction leads to an increase of CBF up to 90-110 bpm: the slight diastole reduction does not promote consistent SV reduction, thus CBF overall grows between 50 and 90 bpm. Only beyond 90-110 bpm, the reduced diastole length makes SV, and therefore CBF, rapidly decrease, as generally expected \cite{Heusch2008}. In terms of myocardial oxygen supply and demand, both CBF and RPP increase with HR up to 90 bpm, albeit with different slopes. Once 90 bpm is exceeded, the oxygen demand still grows with HR, while CBF reaches an optimal value and then decreases. This trend implies that, up to 90 bpm, the coronary perfusion increases trying to supply the increased myocardial oxygen demand. Beyond 90-110 bpm, instead, there is a progressive coronary impairment and a consequent imbalanced oxygen supply-demand ratio.

\smallskip

(ii) irregularity induced by the AF-beating is not absorbed uniformly in the coronary circulation at different HRs. In general, although cv values of the RR beating are kept constant (and equal to 0.24) at all HRs, none of the hemodynamic parameters shows constant cv values as function of HR. In particular, cv values of the hemodynamic parameters exploited for the amplitude evaluation (i.e., $Q_{max,sys}$, $Q_{min}$, and $Q_{max,dia}$) overall increase with HR, but with different behaviours in the range [50,130] bpm: $Q_{max,sys}$ reaches a maximum cv around 90 bpm, $Q_{min}$ has a minimum cv value for 70 bpm, while cv for $Q_{max,dia}$ increases monotonically. Maxima of the LAD flow rate ($Q_{max,sys}$ and $Q_{max,dia}$) present the highest cv increase (for both parameters, cv almost doubles from 50 to 130 bpm), and are the most prone to the AF-induced variability. Also coronary perfusion indexes, such as $SV$ and $CBF$, are deeply influenced by the RR-induced irregularity, showing a monotonic growth of cv with HR.

\noindent The increased coronary hemodynamic variability with HR is well evidenced by the CBF(CPP) scatter plots of Fig. 5, which are much sparser and less predictive for higher HR. Based on the present findings, for $HR\ge90$ bpm CPP cannot be considered a good marker for coronary perfusion in AF. Moreover, the beat-to-beat variability, mainly related to the current RR, is an \emph{a posteriori} proof of the uncorrelated nature of the AF beating and shows how this RR feature is assimilated by the coronary perfusion: coronary blood flow per beat is only consequence of $RR$ and $RR_{-1}$, while it is substantially independent of pre-preceding beats.

The coronary hemodynamics in AF is therefore influenced by the combined effects of both heart rate variations and higher hemodynamic variability. If for the heart rate there is an optimal value (around 90 bpm) which maximizes CBF, the coronary perfusion variability increases monotonically with HR, thus being much more impacting at higher HR. Experimental data in humans demonstrated that the raise in myocardial blood flow accompanying AF may be insufficient to compensate for the increased cardiac workload due to the AF-related rise in heart rate, differently to what happens to the same hearts during atrial pacing at similar ventricular rate \cite{Kochiadakis2002}. In addition to this reduction in coronary blood flow due to the irregularity of the RR intervals, a recent work \cite{Wijesurendra2018} suggests that an underlying microvascular dysfunction, strictly related to left ventricular and left atrial mechanical dysfunction, may play a role in myocardial blood flow impairment in AF patients. In a similar context, the present computational framework focuses on the effects that different ventricular rates exert on the coronary circle during AF. Regardless of baseline clinical features which could act as possible confounders and potential underlying microvascular dysfunction, the present model suggests clinically relevant interpretations. In fact, also here, higher ventricular rates during AF related to worse oxygen demand-supply ratio, indicating that not only the irregularity of the rhythm, but even the ventricular response in AF, may play a significant role in impairing coronary blood flow.

\section{Limitations}

The present model is open-loop and only considers the left heart-arterial tree, so it neglects - similarly for all the HRs - the hemodynamic feedbacks of the systemic baroreceptor mechanisms. Second, the coronary model assumes an autoregulation mechanism related to the intramyocardial pressure and myocardial filling, but does not directly account for the metabolic regulation. Moreover, the presence of other pathologies as well as the assumption of rate control drugs is not taken into account, since the focus is on the pure hemodynamic response to rate and rhythm variations induced by AF. In the end, the 0D-1D model is not able to account for some mechanical properties of both cardiac vessels (such as viscoelasticity of extracellular matrix) and cardiac muscle (such as strain rate), which are important markers of cardiac disease \cite{Huveneers2015}. Given this, however, the fundamental aim of the present work is to use a simplified but powerful computational framework to infer how coronary circle responds to an irregular heartbeat at different mean ventricular rates during AF. The use of standardized conditions, regardless of any other baseline clinical feature that could alter mechanical properties of both cardiac vessels and cardiac muscle, allows us to explore the standalone impact that the irregular AF beating exerts on the coronary circle.

\section{Conclusions}

The present computational study shows that higher ventricular rate during AF exerts a coronary blood flow impairment. In particular, the combined effects of both heart rate variations and higher AF-induced variability leads to the following main outcomes:

(i) waveform and amplitude of the coronary blood flow rate are deeply modified by increasing HR, with a substantial delay of the diastolic peak and a consequent flow rate reduction. The overall increase of mean systolic ($Q_{max,sys}$) and diastolic ($Q_{max,dia}$) flow rate peaks with HR is also accompanied by an increased variability, while negligible effects are encountered for $Q_{min}$ as HR varies;

(ii) as a consequence of waveform variations, CBF presents an optimal value around 90-110 bpm. However, RPP monotonically grows with HR. Thus, exceeding 90-110 bpm in AF, coronary perfusion per beat begins to be impaired and the oxygen supply-demand ratio imbalanced;

(iii) CBF positively correlates with CPP up to 70 bpm, but for higher HR the correlation dramatically drops towards zero and data become sparse. During AF, CPP is no longer a good estimate of the myocardial perfusion for HR higher than 90 bpm.

\section*{Conflicts of interest}

The authors declare that they have no conflicts of interest.

\section*{Acknowledgments}

This study was performed thanks to the support of the "Compagnia di San Paolo" within the project "Progetti di Ricerca di Ateneo – 2016: Cerebral hemodynamics during atrial fibrillation (CSTO 160444)" of the University of Turin, Italy. The funders had no role in study design, data collection and analysis, decision to publish, or preparation of the manuscript.

\end{document}